\documentstyle[12pt,epsf]{article}
\input{epsf}
\newcommand{\N}{{\rm I \hspace{-0.52ex} N}}
\newcommand{\unop}{{\rm 1 \hspace{-0.52ex} I}}
\begin{document}
\setlength\textheight{8.75in}
\newcommand{\be}{\begin{equation}}
\newcommand{\ee}{\end{equation}}
\title{ Extended Jaynes-Cummings models
and (quasi)-exact solvability
}
\author{{\large Yves Brihaye\footnote{brihaye@umh.ac.be}
and Ancilla Nininahazwe
\footnote{nininaha@yahoo.fr}}\\
{\small Facult\'e des sciences, Universit\'e de Mons-Hainaut}\\
{\small B-7000 Mons, Belgium}}
\date{\today}
\maketitle
\thispagestyle{empty}
\begin{abstract}
		The original Jaynes-Cummings model is 
                described by a Hamiltonian which is  hermitian and
                exactly solvable. Here we extend this model by 
           several types of interactions leading to a non hermitian
                operator which doesn't satisfy the physical condition 
                of space-time reflection symmetry (PT symmetry). 
                The new Hamiltonians are either exactly solvable
                admitting  an entirely  real spectrum or quasi exactly
                solvable with a real algebraic part of their spectrum.
\end{abstract}
\medskip
\medskip
\newpage
\section{Introduction}
Several new theoretical aspects of quantum mechanics have been developped
in the last years. In series of papers (see e.g. \cite{bender1,bender2}
and \cite{bender3} for a recent review) it is shown that the traditional
self adjointness requirement of the Hamiltonian operator is not a necessary
condition to guarantee a real spectrum and that the weaker condition
of PT-invariance of the Hamiltonian is sufficient for that purpose.
An alternative possibility for an operator to admit a real spectrum is
also developed in \cite{mostafa}.
It is the notion of pseudo-hermiticity.
Following the ideas of \cite{mostafa},
we remind here that a Hamiltonian
        is called $\eta$ pseudo-hermitian if it satisfies the
        relation $\eta H \eta^{-1} = H^{\dagger}$, where $\eta$ 
        denotes a linear hermitian operator.
        It is this new notion (i.e pseudo-hermiticity property)
        of non hermitian Hamiltonians which explains the reality 
        of their energy spectrum. This important property has further been
        considered in Refs.\cite{Mandal,Pijush}.

Another direction of development of quantum mechanics is the notion
of quasi exact solvability \cite{turbiner,ushveridze}. It provides techniques
to construct linear operators preserving a {\it finite dimensional} 
subspace ${\cal V}$ of the Hilbert space. Accordingly, the so called
Quasi Exactly Solvable operators, once restricted on ${\cal V}$ can
be diagonalized by means of algebraic methods. The QES property is
strongly connected to finite dimensional representation of Lie or graded
Lie algebras \cite{turbiner,shiftur,bk}. Amongst many models used to
describe quantum properties of physical systems, the Jaynes-Cummings
model play an important role \cite{Gerry, Ray, knight,nathalie}.
It describes, in a simple way the interaction of photons with a spin-1/2
particle.
From the
mathematical point of view, the Jaynes-Cummings model
is described by a self-adjoint operator and it is completely solvable
in a sense that the entire spectrum can be computed algebraically.

	The purpose of this paper is to consider operators generalizing
        the Jaynes-Cummings Hamiltonians
        which are neither self-adjoint nor PT-invariant
        but which are pseudo-hermitian with respect to two different
        operators.
        In particular, from the original Jaynes-Cummings model
        (JCM in the following),
        we construct an extended one by adding a  polynomial
        of the form $P(a^{\dagger},a)$ ($ a^{\dagger}, a$ are the usual
        creation and annihilation operators)
        of degree $d \geq 2$ in the diagonal part of the hamiltonian.
        Some particular choices of $P$ are constructed in such a way that
         the resulting operator becomes QES.
        The non-diagonal interaction
        part is also modified in such a way that (i) multiple
        photon exchanges are allowed and (ii) the
        full operator can be hermitian or pseudo-hermitian.

Here is the plan of the paper.
In  section 2, we revisit the  Hamiltonian
considered  in Ref.\cite{Mandal} and express it in terms of differential
operator of a real variable $x$.
This reveals its exact solvability in terms of differential operators
acting on sets of polynomials of appropriate degrees in $x$.
In Sect. 3 we propose a family of operators which generalize
the original JC Hamiltonian in several respects. The (pseudo)-hermiticity
of these operators are analysed and the spectra and the eigenvectors
are computed in details for a few of them. The differences in the
spectrum corresponding to Hermitian and pseudo-Hermitian are pointed out.
In particular, the energy eigenvalues are entirely real in spite of the fact
that  they are associated to a non hermitian and
         non $PT$-invariant Hamiltonian. The reality
         of those eigenvalues is ensured by
         the pseudo-hermiticity of the Hamiltonians.
         The section 4 is devoted to
         QES extensions of the JCM. These are constructed in such a way
        that, both, one-photon and two-photons exchange terms coexist in the
        non-diagonal interacting terms.
         By construction, these
          new models preserve  finite
          dimensional vector spaces of the Hilbert spaces
          ,the algebraic part of the spectrum is computed in Sect 5.
 Further properties of these new types of QES operators, say $H_T$,
 can be discussed.
 Namely, following the ideas of \cite{benderdunne}
 we show in Sect. 6 that the solutions of the spectral equation
 $H_T \psi = E \psi$
 for generic values of $E$ lead to  new types
 of recurence relations.
 The relations between $H_T$ and specific graded algebras are
 pointed out in Sect 7.
 Finally, the section 8 is kept
          for concluding remarks.

\section{Exactly solvable pseudo-hermitian Hamiltonian}
In this section we consider the Hamiltonian
describing a system of a spin-$\frac{1}{2}$ particle
in the external magnetic field, $\vec B $ which couples
to a harmonic oscillator through some nonhermitian interaction 
\cite{Mandal}
\be
\label{Ham}
 H = \mu\vec \sigma\cdot\vec B + \hbar \omega a^{\dagger}a
 + \rho(\sigma_+a - \sigma_-a^{\dagger}).
\ee
Here $\vec\sigma$ denotes Pauli matrices, $\rho$ is some 
arbitrary real parameter and 
$\sigma_{\pm}\equiv \frac{1}{2}[\sigma_x\pm i\sigma_y]$. 
$\sigma_+$ and $\sigma_-$ can be expressed in matrix form
\be
\sigma_+ =\left(\matrix{0&1\cr
                        0&0\cr}\right)\ \ , \ \ \sigma_- =\left(\matrix{0&0\cr
                                                                   1&0\cr}\right).
\ee
Our purpose is to relate the  Hamiltonian above to an appropriate
differential operator preserving a family of spaces of polynomials
in the variable $x$, following the ideas
 of exactly and quasi-exactly solvable operators
\cite{turbiner}.
With this aim, we use the usual creation and annihilation 
operators respectively
$a^{\dagger}$ and $a$ which are defined as follows
\be
\label{creation}
a^{\dagger} = \frac{p + im\omega x}{\sqrt {2m\omega\hbar}} \ \ , \ \ a= \frac{p - im\omega x}{\sqrt{ 2m\omega\hbar}},
\ee
where $p = -i\frac{d}{dx}$.
The external magnetic field is chosen in $z$-direction (i.e $\vec B = B_0\vec z$) in order to reduce the Hamiltonian defined in Eq.(\ref{Ham}) and it has the form
\be
\label{reduced}
H = \frac{\epsilon}{2}\sigma_z + \hbar\omega a^{\dagger}a + \rho(\sigma_+a - \sigma_-a^{\dagger}),
\ee
where $\epsilon = 2\mu B_0$.
As $\sigma_{\pm}^{\dagger} = \sigma_{\mp}$, it is pointed out that this Hamiltonian is not hermitian
\begin{eqnarray}
&H^{\dagger} &= \frac{\epsilon}{2}\sigma_z + \hbar\omega a^{\dagger}a - \rho(\sigma_+a - \sigma_-a^{\dagger}),\nonumber\\
&   &\not = H.
\end{eqnarray}
Thus as,
\begin{eqnarray}
& PT H (PT)^{-1}&= -\frac{\epsilon}{2}\sigma_z + \hbar\omega a^{\dagger}a + \rho(\sigma_+a^{\dagger} - \sigma_-a),\nonumber\\
&   &\not = H,
\end{eqnarray}
one can see that the Hamiltonian (\ref{Ham}) 
is not PT symmetric i.e $H \not = H^{PT}$ \cite{bender1}.

The next step is to write $H$ in terms of differential
operators(i.e $ p= -i\frac{d}{dx}$) and of variable $x$.
The purpose of these transformations is to reveal the
exact solvability of the operator $H$ by using the 
quasi-exactly solvable (QES) technique as has been 
considered in Ref.\cite{nathalie}.
Replacing the operators $a^{\dagger}$ and $a$ by their expressions(as given in Eq.(\ref{creation}))in the Eq.(\ref{reduced}), the Hamiltonian of the model is written now as follows 
\be
H = \frac{\epsilon}{2}\sigma_z  +  \frac{p^2 - m\omega + m^2\omega^2x^2}{2m} + \rho \frac{[\sigma_+(p-im\omega x)- \sigma_-(p + im\omega x)]}{\sqrt {2m\omega\hbar}}
\ee
In order to reveal the solvability of the above operator $H$, we first perform the standard (often called " gauge") transformation
\be
\tilde H = R^{-1}H R \ \ , \ \ R = exp(-\frac{m\omega x^2}{2}).
\ee
After some algebra, the new Hamiltonian $\tilde H$ is obtained and is given by
\begin{eqnarray}
&\tilde H&= \frac{\epsilon}{2}\sigma_z  - \frac{1}{2m}\frac{d^2}{dx^2} + \omega x\frac{d}{dx} + \rho \frac{[\sigma_+ p - \sigma_- (p + 2im\omega x)]}{\sqrt{2m\omega\hbar}}\nonumber\\
&        &=\frac{\epsilon}{2}\sigma_z  + \frac{p^2}{2m} + \omega x\frac{d}{dx} + \rho \frac{[\sigma_+ p - \sigma_- (p + 2im\omega x)]}{\sqrt{2m\omega\hbar}}
\end{eqnarray}
Replacing $\sigma_z$, $\sigma_+$ and $\sigma_-$ by their matrix form, the final form of the Hamiltonian $\tilde H$ reads
\begin{eqnarray}
& \tilde H &= \frac{\epsilon}{2}\left(\matrix{1&0\cr
                                             0&-1\cr}\right) + \left(\matrix{\frac{p^2}{2m} + \omega x\frac{d}{dx}&0\cr
                        0&\frac{p^2}{2m} + \omega x\frac{d}{dx}\cr}\right) - \rho \left(\matrix{0&-\frac{p}{\sqrt{2m\omega\hbar}}\cr
                                             \frac{p + 2im\omega x}{\sqrt{2m\omega\hbar}}&-0\cr}\right)\nonumber\\
&          &= \left(\matrix{\frac{p^2}{2m} + \omega x\frac{d}{dx} + \frac{\epsilon}{2}&\rho \frac{p}{\sqrt{2m\omega\hbar}}\cr
                                             -\rho\frac{p + 2im\omega x}{\sqrt{2m\omega\hbar}}&\frac{p^2}{2m} + \omega x\frac{d}{dx} - \frac{\epsilon}{2}\cr}\right).                                             
\end{eqnarray}
Then, the operator $\tilde H$ is typically QES because it preserves a finite dimensional vector spaces of polynomials namely ${\cal V}_n = (P_{n-1}(x), P_n(x))^t$ with $n\in \N$. Moreover $\tilde H$ is exactly solvable because $n$ does not have to be fixed (it can be any nonnegative integer).

Note that the above Hamiltonian $\tilde H$ is not 
invariant under simultaneous parity operator(P) and time reversal 
(T)reflection (i.e respectively $x\rightarrow -x$ and $i\rightarrow -i$)
\cite{bender1}.
Even if the operator $\tilde H$(therefore $H$)is nonhermitian and not PT invariant, it was pointed out that its spectrum is real. The reality of the eigenvalues of $H$ is a consequence of the unbroken $P\sigma_z$(i.e combined parity operator $P$ and Pauli matrice $\sigma_z$)invariance of $H$ (i.e $[H,P\sigma_z] = 0$). In other words, the spectrum is real because $H$ is pseudo-hermitian with respect to $\sigma_z$(i.e $\sigma_z H\sigma_z^{-1}= H^{\dagger}$) and also to the parity operator $P$ (i.e $PHP^{-1} = H^{\dagger}$) 
\cite{mostafa,Mandal,Pijush}.
We would like to mention that it is not necessary 
to calculate the energy eigenvalues and their 
corresponding eigenvectors of $H$ because they have been determined 
in \cite{Mandal}. In the following section, we will construct 
the spectrum of the generalized Hamiltonian of the one given by Eq.(\ref{Ham}).
\section{Family of exactly solvable Hamiltonians}
The original JCM is defined by the Hamiltonian 
\be
\label{jcm}
H = \frac{\epsilon}{2}\sigma_3 + \hbar\omega a^{\dagger}a
+ \rho( a\sigma_+ +  a^{\dagger}\sigma_-),
\ee
where $\rho $ is a real parameter(i.e it is a real coupling constant). 
Note here that the Hamiltonian $H$ is  hermitian.

In the next, we consider an extension of the above 
JCM Hamiltonian in the form
\be
\label{ht}
H = \frac{\epsilon}{2}\sigma_3 + \hbar\omega a^{\dagger}a 
+ P(a^{\dagger}a ) + \rho(a^k\sigma_+  + \phi (a^{\dagger})^k\sigma_-),
\ee
where $\phi = \pm 1$
and $P(a^{\dagger}a )$ denotes a polynomial of degree 
$d\geq 2$, $k$ is an integer $\geq 1$ and $\rho$ is an 
arbitrary real parameter. In fact, the above Hamiltonian 
is nonhermitian( i.e for $\phi = -1$ ) and not PT invariant but it satisfies 
the pseudo-hermiticity with the operators $P$ 
(operator of parity) and $\sigma_3$
(Pauli matrice ). considering  $\phi = +1$ the Hamiltonian given 
by the Eq.(\ref{ht})
becomes hermitian. Both for these cases, it can be 
easily observed that the energy spectrum is entirely real. 
Thus, notice that the above Hamiltonian (\ref{ht}) is a generalization 
of the Hamiltonians given by the Eqs.(\ref{Ham}) and (\ref{jcm}).
The matrix form of $H$ reads
\be
\label{particular}
\left(\matrix{\hbar\omega a^{\dagger}a 
+ P(a^{\dagger}a) + \frac{\epsilon}{2}&\rho a^k\cr
                        \phi\rho(a^{\dagger})^k&\hbar\omega a^{\dagger}a 
                        + P(a^{\dagger}a) - \frac{\epsilon}{2}\cr}\right)
\ee
which can be easily checked to preserve the vector spaces
\be
{\cal V}_n = span\Biggl\{\left(\matrix{\mid n \rangle\cr
                        0\cr}\right),\left(\matrix{0\cr
                        \mid n+k\rangle\cr}\right)\Biggr\} \ \ , \ \ n\in \N.
\ee
It means that the action of the operator $H$ on the vectors
states $\left(\matrix{\mid n \rangle\cr
                        0\cr}\right)$ 
                        and $\left(\matrix{0\cr
                        \mid n+k\rangle\cr}\right)$
can expressed  as linear combinations of these same states. 
Here, we are allowed to conclude that $H$ is 
exactly solvable because it preserves the vector space 
${\cal V}_n$ for any integer $n$.

The next step is to find the energy eigenvalues and their 
corresponding eigenvectors of the Hamiltonian $H$ for $\phi = -1$ 
 and for $\phi = +1$. For this purpose we recall the following 
identities\cite{Mandal}
\begin{eqnarray} 
&a^{\dagger}a\mid n,\frac{1}{2}m_s\rangle &= n\mid n,\frac{1}{2}m_s\rangle,\nonumber\\
&\sigma_3\mid n,\frac{1}{2}m_s\rangle &= m_s\mid n,\frac{1}{2}m_s\rangle,\nonumber\\
&\sigma_+\mid n,\frac{1}{2}\rangle &= 0 \ \ ; \ \ \sigma_+\mid n,-\frac{1}{2}\rangle 
= \mid n,\frac{1}{2}\rangle,\nonumber\\
&\sigma_-\mid n,-\frac{1}{2}\rangle &= 0 \ \ ; \ \ \sigma_-\mid n,\frac{1}{2}\rangle 
= \mid n,-\frac{1}{2}\rangle,
\end{eqnarray} 
with $n$ and $m_s = \pm 1$ are respectively the eigenvalues of the number 
operator $a^{\dagger}a$ and the operator $\sigma_3$. It is readily seen 
that the state $\mid 0,-\frac{1}{2}\rangle$ is a ground state of the 
operator $H$(i.e it is constructed by the lowest values of $n$ and $m_s $ 
which are respectively $0$ and $-1$). We have now to consider the 
action of $H$ to the state $\mid 0,-\frac{1}{2}\rangle$ in order to find 
its associated eigenvalue
\begin{eqnarray}
& H \mid 0,-\frac{1}{2}\rangle &= \frac{\epsilon}{2}\sigma_3\mid 0,-\frac{1}{2}\rangle + \hbar\omega a^{\dagger}a\mid 0,-\frac{1}{2}\rangle + P(a^{\dagger}a )\mid 0,-\frac{1}{2}\rangle \nonumber\\
&   & + \rho a^k\sigma_+\mid 0,-\frac{1}{2}\rangle 
 +  \phi \rho (a^{\dagger})^k\sigma_-\mid 0,-\frac{1}{2}\rangle,\nonumber\\
&    &= \frac{\epsilon}{2}\sigma_3\mid 0,-\frac{1}{2}\rangle,\nonumber\\
&    &= -\frac{\epsilon}{2}\mid 0,-\frac{1}{2}\rangle.
\end{eqnarray}
It is proved now that $-\frac{\epsilon}{2}$ is the eigenvalue of the ground state $\mid 0,-\frac{1}{2}\rangle$. 
It is easily understood that the next state $\mid 0,\frac{1}{2}\rangle$ is not an eigenstate alone of 
the Hamiltonian $H$ because applying this operator to this state, we obtain a linear combination 
of two states $\mid 0,\frac{1}{2}\rangle$ and $\mid k,-\frac{1}{2}\rangle$,
\be
H \mid 0,\frac{1}{2}\rangle = \frac{\epsilon}{2}\mid 0,
\frac{1}{2}\rangle \pm \rho\sqrt {k!}\mid k,-\frac{1}{2}\rangle.
\ee
The state $\mid k,-\frac{1}{2}\rangle$ under the action of $H$ 
leads to a linear combination also of
 two above states 
\be
H \mid k,-\frac{1}{2}\rangle = (\hbar\omega k + P(k)-
\frac{\epsilon}{2})\mid k,-\frac{1}{2}\rangle + 
\rho\sqrt {k!}\mid 0,\frac{1}{2}\rangle.
\ee 
The excited states $\mid 0,\frac{1}{2}\rangle$ and $\mid k,-\frac{1}{2}\rangle$ span an invariant 
subspace in the space of states so that the Hamiltonian matrix is written as follows
\be
\label{mat}
H_k = \left(\matrix{\frac{\epsilon}{2}&\rho\sqrt {k!}\cr
                        \phi \rho\sqrt {k!}&\hbar\omega k + P(k)-\frac{\epsilon}{2}\cr}\right)
\ee
In particular, note that for $k=1$, $P(k)=0$(i.e $P(k)= k^d$ \ , \ $d \geq 2$) and 
considering $\phi = -1$, $H_k$ becomes the matrix $H_1$ constructed in \cite{Mandal}.
In order to find the eigenvalues of the Hamiltonian matrix(\ref{mat}), we have to solve 
the following usual equation(i.e characteristic polynomial)
\begin{eqnarray}
&det(H_k - \lambda \unop) &= 0,\nonumber\\
&\left(\matrix{\frac{\epsilon}{2}-\lambda&\rho\sqrt {k!}\cr
                        \phi \rho\sqrt {k!}&\hbar\omega k + P(k)
                        -\frac{\epsilon}{2} - \lambda\cr}\right) 
                        &= 0,\nonumber\\
&4\lambda^2 - 4(\hbar\omega k + P(k))\lambda 
+ 2(\hbar\omega k + P(k))\epsilon - \epsilon^2 +  \phi 4k!\rho^2 &=0.
\end{eqnarray}
After some algebra, the energy eigenvalues(i.e square-roots of the above equation in $\lambda$) 
of $H_k$ are 
\begin{eqnarray}
\label{eigen}
&\lambda_k^I &= \frac{\hbar\omega k + P(k) +
\sqrt{(\hbar\omega k + P(k) - \epsilon)^2 + \phi 4k!\rho^2}}{2}, \nonumber\\
&\lambda_k^{II}&= \frac{\hbar\omega k + P(k) -
\sqrt{(\hbar\omega k + P(k) - \epsilon)^2  + \phi 4k!\rho^2}}{2}.
\end{eqnarray}
It is easily checked that for $k=1$, $P(k)=0$ and for $\phi = -1$), 
we obtain the eigenvalues $\lambda_1^{I,II}$ determined in\cite{Mandal}. These are the energy eigenvalues of the Hamiltonian (\ref{Ham}).
The next step now is to calculate the associated eigenvectors of the above eigenvalues $\lambda_k^{I,II}$. Here, we propose to consider two cases : the first case for $\phi = -1$ and the second one for $\phi = +1$.

\subsection{The case  $\phi = -1$}
Considering $\phi = -1$, the eigenvalues (\ref{eigen}) are given by
\begin{eqnarray}
&\lambda_k^I &= \frac{\hbar\omega k + P(k) + \sqrt{(\hbar\omega k + P(k) - \epsilon)^2 - 4k!\rho^2}}{2}, \nonumber\\
&\lambda_k^{II} &= \frac{\hbar\omega k + P(k) - \sqrt{(\hbar\omega k + P(k) - \epsilon)^2 - 4k!\rho^2}}{2}.
\end{eqnarray}
For the sake simplicity, we can impose $P(k)=0$ and the eigenvalues $\lambda_k^{I,II}$ have the form
\begin{eqnarray}
&\lambda_k^I &= \frac{\hbar\omega k + \sqrt{(\hbar\omega k - \epsilon)^2 - 4k!\rho^2}}{2}, \nonumber\\
&\lambda_k^{II} &= \frac{\hbar\omega k - \sqrt{(\hbar\omega k - \epsilon)^2 - 4k!\rho^2}}{2}.
\end{eqnarray}       
The following relations are considered as in \cite{Mandal} 
\begin{eqnarray}
\label{sin}
&\vert \hbar\omega k - \epsilon \vert &\geq 2\rho\sqrt{k!},\nonumber\\
&2\rho\sqrt{k!} &= (\hbar\omega k - \epsilon)\sin\theta_k
\end{eqnarray} 
and the Hamiltonian matrix given by (\ref{mat}) reads
\begin{eqnarray}
&H_k &= \left(\matrix{\frac{\epsilon}{2}&\rho\sqrt {k!}\cr
                        -\rho\sqrt {k!}&\hbar\omega k -\frac{\epsilon}{2}\cr}\right),\nonumber\\
&     &= \left(\matrix{\frac{\epsilon}{2}&\frac{1}{2}(\hbar\omega k - \epsilon)\sin\theta_k\cr
                        -\frac{1}{2}(\hbar\omega k - \epsilon)\sin\theta_k&\hbar\omega k -\frac{\epsilon}{2}\cr}\right).   
\end{eqnarray} 
Taking account of the following equation
\be
\left(\matrix{\frac{\epsilon}{2}&\frac{1}{2}(\hbar\omega k - \epsilon)\sin\theta_k\cr
                        -\frac{1}{2}(\hbar\omega k - \epsilon)\sin\theta_k&\hbar\omega k -\frac{\epsilon}{2}\cr}\right)\left(\matrix{A\cr
                                           B\cr}\right) = \lambda_k^{I,II}\left(\matrix{A\cr
                                           B\cr}\right),
\ee
the associated eigenvectors of $\lambda_k^{I,II}$ are determined
\begin{eqnarray}
\label{state0}
&\mid \psi_k^I\rangle  &= \sin\frac{\theta_k }{2}\mid 0, \frac{1}{2}\rangle + \cos\frac{\theta_k }{2}\mid k, -\frac{1}{2}\rangle,\nonumber\\
&              &for \ \lambda_k^I = \frac{\hbar\omega k}{2}(1 + \cos \theta_k) - \frac{\epsilon}{2}\cos \theta_k,
\end{eqnarray}
with  $A = \sin\frac{\theta_k }{2}$ and $B = \cos\frac{\theta_k }{2}$.

\begin{eqnarray}
\label{state1}
&\mid \psi_k^{II}\rangle & = \cos\frac{\theta_k }{2}\mid 0, \frac{1}{2}\rangle + \sin\frac{\theta_k }{2}\mid k, -\frac{1}{2}\rangle,\nonumber\\
&      &for\ \lambda_k^{II} = \frac{\hbar\omega k}{2}(1 - \cos {\theta_k}) + \frac{\epsilon}{2}\cos {\theta_k},
\end{eqnarray}
with $A = \cos\frac{\theta_k }{2}$ and $B = \sin\frac{\theta_k }{2}$.

In particular, for $k=1$, it is easily checked that $\psi_k^I$ and $\psi_k^{II}$ 
become respectively $\psi_1^I$ and $\psi_1^{II}$ which were determined in \cite{Mandal}.

\subsection{The case $\phi=+1$}
Taking account of $\phi=+1$ and imposing $P(k)=0$ , 
the eigenvalues (\ref{eigen}) read
\begin{eqnarray}
\label{sinh}
&\lambda_k^I &= \frac{\hbar\omega k + \sqrt{(\hbar\omega k 
- \epsilon)^2 + 4k!\rho^2}}{2}, \nonumber\\
&\lambda_k^{II} &= \frac{\hbar\omega k  - \sqrt{(\hbar\omega k 
- \epsilon)^2 + 4k!\rho^2}}{2}.
\end{eqnarray}
The relations considered in Eq.(\ref{sin}) become
\begin{eqnarray}
&\vert \hbar\omega k - \epsilon \vert &\geq 2\rho\sqrt{k!},\nonumber\\
&2\rho\sqrt{k!} &= (\hbar\omega k - \epsilon)\sinh\theta_k.
\end{eqnarray}                        
Following the same method used in the previous case, the eigenvectors associated 
to above eigenvalues (\ref{sinh}) are written as follows
\begin{eqnarray}
\label{state2}
&\mid \psi_k^I\rangle  &= \sinh\frac{\theta_k }{2}\mid 0, \frac{1}{2}\rangle 
+ \cosh\frac{\theta_k }{2}\mid k, -\frac{1}{2}\rangle,\nonumber\\
&              &for \ \lambda_k^I = \frac{\hbar\omega k}{2}
(1 + \cosh \theta_k) - \frac{\epsilon}{2}\cosh \theta_k,\nonumber\\
&\mid \psi_k^{II}\rangle & = \cosh\frac{\theta_k }{2}\mid 0, \frac{1}{2}\rangle 
- \sinh\frac{\theta_k }{2}\mid k, -\frac{1}{2}\rangle,\nonumber\\
&      &for\ \lambda_k^{II} = 
\frac{\hbar\omega k}{2}(1 - \cosh {\theta_k}) 
+ \frac{\epsilon}{2}\cosh {\theta_k},
\end{eqnarray} 

For $H \not= H^{\dagger}$ (i.e for $\phi = -1$), it may be easily observed that two
 states given in (\ref{state0}) and (\ref{state1}) are not orthogonal to each other. 
 But one can prove that the states given by Eq.(\ref{state2}) (i.e for $\phi = +1$, 
 $H = H^{\dagger}$) are orthogonal.This property is a consequence of the hermiticity of $H$.
  In order to find the next excited states, one has to consider the next 
 invariant  subspace which is spanned by the vectors $\mid 1, \frac{1}{2}\rangle$ and $\mid k+1, -\frac{1}{2}\rangle$. 
 The eigenvalues and eigenvectors for this doublet can be determined following the same method
 used previously.

\subsection{The excited states}
The next step is to generalize the previous results to the invariant subspace which 
is spanned by the vectors   $\mid n, \frac{1}{2}\rangle$ and $\mid n+k, -\frac{1}{2}\rangle$. 
Following the same technique used in the previous section and after some algebra, 
the Hamiltonian matrix for the above doublet is written as,
\be                                
H_{n+k} = \left(\matrix{\hbar\omega n + P(n) + \frac{\epsilon}{2}&\rho\sqrt {n+1} \dots \sqrt{n+k}\cr
                        \phi \rho\sqrt {n+1} \dots \sqrt{n+k} &\hbar\omega(n+k) + P(n+k)-\frac{\epsilon}{2}\cr}\right)
\ee
For the sake simplicity, we impose $P(n) = P(n+k) = 0$ and $H_{n+k}$ is of the form
\be
H_{n+k} = \left(\matrix{\hbar\omega n + \frac{\epsilon}{2}&\rho\sqrt {n+1} \dots \sqrt{n+k}\cr
                        \phi \rho\sqrt {n+1} \dots \sqrt{n+k} &\hbar\omega(n+k)-\frac{\epsilon}{2}\cr}\right)
\ee
and its eigenvalues are
\begin{eqnarray}
\label{lam}
\lambda_{n+k}^I = \frac{\hbar\omega(2n + k) + 
\sqrt{(\hbar\omega k  - \epsilon)^2 + \phi 4\rho^2(n+1)\dots(n+k)}}{2}, 
\nonumber\\
\lambda_{n+k}^{II} = \frac{\hbar\omega(2n + k) -
\sqrt{(\hbar\omega k  - \epsilon)^2 + \phi 4 \rho^2(n+1)\dots(n+k)}}{2},
\end{eqnarray}
In particular, putting $k=1$ and $\phi=-1$ only in (\ref{lam}), 
the above eigenvalues become the eigenvalues $\lambda_{n+1}^{I,II} $ associated to the operator
 $H$ given by the Eq.(\ref{Ham}). These eigenvalues were determined in \cite{Mandal}.

Now putting $2\rho\sqrt {n+1} \dots \sqrt{n+k} = (\hbar\omega k  - \epsilon)\sin\theta_{n+k}$ and
$2\rho\sqrt {n+1} \dots \sqrt{n+k} = (\hbar\omega k  - \epsilon)\sinh\theta_{n+k}$ in Eq.(\ref{lam})
respectively for $\phi=-1$ and for $\phi=+1$, we find the eigenvectors corresponding to the doublet 
$\mid n, \frac{1}{2}\rangle$ and $\mid n+k, -\frac{1}{2}\rangle$.

First considering $\phi=-1$, the eigenvectors associated to this doublet are
\begin{eqnarray}
\label{cos}
&\mid \psi_{n+k}^I\rangle  &= \sin\frac{\theta_{n+k} }{2}\mid n, \frac{1}{2}\rangle + \cos\frac{\theta_{n+k} }{2}\mid n+k, -\frac{1}{2}\rangle,\nonumber\\
&              &for \ \lambda_{n+k}^I =  \hbar\omega n + \frac{\hbar\omega k}{2}(1 + \cos \theta_{n+k}) - \frac{\epsilon}{2}\cos \theta_{n+k},\nonumber\\
&\mid \psi_{n+k}^{II}\rangle  &= \cos\frac{\theta_{n+k} }{2}\mid n, \frac{1}{2}\rangle + \sin\frac{\theta_{n+k} }{2}\mid n+k, -\frac{1}{2}\rangle,\nonumber\\
&              &for \ \lambda_{n+k}^I =  \hbar\omega n + \frac{\hbar\omega k}{2}(1 -\cos \theta_{n+k}) + \frac{\epsilon}{2}\cos \theta_{n+k}.
\end{eqnarray}

Finally considering $\phi=+1$ for the Eq.(\ref{lam}), the eigenvectors 
for the doublet $\mid n, \frac{1}{2}\rangle$ and $\mid n+k, -\frac{1}{2}\rangle$ are of the form
\begin{eqnarray}
&\mid \psi_{n+k}^I\rangle  &= \sinh\frac{\theta_{n+k} }{2}\mid n, \frac{1}{2}\rangle + 
\cosh\frac{\theta_{n+k} }{2}\mid n+k, -\frac{1}{2}\rangle,\nonumber\\
&              &for \ \lambda_{n+k}^I =  \hbar\omega n + \frac{\hbar\omega k}
{2}(1 + \cosh \theta_{n+k}) - \frac{\epsilon}{2}\cosh \theta_{n+k},\nonumber\\
&\mid \psi_{n+k}^{II}\rangle  &= \cosh\frac{\theta_{n+k} }{2}\mid n, \frac{1}{2}\rangle  - \sinh\frac{\theta_{n+k} }{2}\mid n+k, -\frac{1}{2}\rangle,\nonumber\\
&              &for \ \lambda_{n+k}^I =  \hbar\omega n + \frac{\hbar\omega k}{2}(1 -\cosh\theta_{n+k}) + \frac{\epsilon}{2}\cosh\theta_{n+k}.
\end{eqnarray}
Note that all the discussions considered in the previous section are confirmed by these generalized results.
\section{Quasi-exactly solvable Hamiltonians}
In this section let us consider an
extension of the Jaynes-Cummings Hamiltonian
which includes two-photon interaction
\be
H_2 = \frac{\epsilon}{2}\sigma_3 + \hbar\omega a^{\dagger}a 
+ \rho(\sigma_+a^2+ \sigma_-{a^{\dagger}}^2)
\ee
The matrix form of the above Hamiltonian reads
\be
H_2 = \left(\matrix{\hbar\omega a^{\dagger}a + \frac{\epsilon}{2}&\rho a^2\cr
                        \rho(a^{\dagger})^2&\hbar\omega a^{\dagger}a  
                        - \frac{\epsilon}{2}\cr}\right).
\ee
It is clear that this Hamiltonian $H$
is similar of the one reported in
Ref.\cite{Gerry} and is also a particular case
of the Hamiltonian given in Eq.(\ref{particular})
(i.e if $k=2, P(a^{\dagger}a)= 0$) and one can prove
easily its exact solvability.
However, if one would like to construct an  JC-type Hamiltonian
including both a one-photon and a two-photon interaction, the
above Hamiltonian should be modified as follows
\be
H_{12} = \left(\matrix{\hbar\omega a^{\dagger}a + \frac{\epsilon}{2}&
\rho a^2 + \rho_1 a \cr
\rho(a^{\dagger})^2 + \hat \rho_1 a^{\dagger}
&\hbar\omega a^{\dagger}a  - \frac{\epsilon}{2}\cr}\right).
\ee
where $\rho, \rho_1 , \hat \rho_1$ are, a priori, arbitrary constants.

Unfortunately,
the corresponding operator $H_{12}$ is not anylonger exactly solvable.
Indeed, it is easy to show that it fails to admit any
finite dimensional invariant vector spaces.
Accordingly, it is impossible (to our knowledge)
to find its energy spectrum by algebraic methods.

In order to restaure, at least partly, a certain algebraic
solvability of $H_{12}$, one
 can attemp to supplement the  Hamitonian $H_{12}$
 with an appropriate  interation term.
After some algebra, one can convince oneself that adding  an interaction
 term of the form
\be
      H_{I} =  \frac{1}{n}\left(\matrix{0&\rho_1aa^{\dagger}a\cr
                         \hat \rho_1a^{\dagger}aa^{\dagger}
                         &0\cr}\right)
\ee
leads to a new Hamiltonian $H_T = H_{12}+ H_{I}$
which is
quasi-exactly solvable, as we will now demonstrate.

Assuming  $n$ to be an integer and redefining
$c \equiv-\frac{\rho_1}{n}$, $\hat c\equiv-\frac{\hat {\rho_1}}{n}$,
the operator $H_T$ reads
\be
\label{operator}
H_T = \left(\matrix{\hbar\omega a^{\dagger}a
+ \frac{\epsilon}{2}&\rho a^2 + c a(a^{\dagger}a-n)\cr
                        \phi \rho(a^{\dagger})^2
+ \hat c(a^{\dagger}a-n)a^{\dagger}&\hbar\omega a^{\dagger}a
- \frac{\epsilon}{2}\cr}\right),
\ee
where that $a^{\dagger}$ and  $a$ are respectively
the usual creation and annihilation operator
and $\epsilon$ is chosen as previously according to
$\epsilon = 2\mu B_0$.

The main idea now is to reveal that the above operator
$H_T$ is quasi-exactly solvable(QES). In this purpose we construct a
finite dimensional vector space which is invariant
under the action of $H_T$.
Let us apply now the Hamiltonian $H$ to the states
$\left(\matrix{\mid N \rangle\cr
                        0\cr}\right)$
                        and $\left(\matrix{0\cr\mid M \rangle\cr}\right)$
                        with $N,M \in \N$ as follows

{\tiny
\be
H_T\left(\matrix{\mid N\rangle\cr
\mid M\rangle\cr}\right)=\left(\matrix{(\hbar\omega N+
\frac{\epsilon}{2})\mid N\rangle+\rho \sqrt{M(M-1)}\mid M-2\rangle+
c\sqrt M (M-n)\mid M-1\rangle\cr\phi\rho\sqrt{(N+1)(N+2)}\mid N+2\rangle+
 \hat c\sqrt{N+1}(N+1-n)\mid N+1\rangle
 +(\hbar\omega M-\frac{\epsilon}{2})\mid M\rangle\cr}\right).
\ee
}
In order to be in agreement with the invariance of the two vectors states 
$\left(\matrix{\mid N \rangle\cr 0\cr}\right)$ and $\left(\matrix{0\cr\mid M \rangle \cr}\right)$ 
under the action of the Hamiltonian $H_T$, we have to impose the value of the integer $n$
 according to $n = M = N+2$ (i.e $N = M-2$).
Taking account of the above fixed value of $n$, we obtain

{\tiny
\be
\label{qes}
H_T \left
(\matrix{\mid N\rangle\cr
\mid M\rangle\cr}\right
)=\left(\matrix{
\biggl
[(\hbar\omega N+
\frac{\epsilon}{2})+\rho\sqrt{(N+2)(N+1)}
\biggr
]\mid N\rangle
\cr
\biggl
[\hbar\omega(N+2)-\frac{\epsilon}{2} + \phi \rho \sqrt{(N+1)(N+2)}
\biggr
]\mid N+2\rangle-\hat c\sqrt{N+1}\mid N+1\rangle\cr}\right).
\ee
}
Finally the Hamiltonian $H_T$ is of the new form
\be
H_T = \left(\matrix{\hbar\omega a^{\dagger}a + 
\frac{\epsilon}{2}&\rho a^2 + c a(a^{\dagger}a-(N+2))\cr
                        \pm\rho(a^{\dagger})^2
                        + \hat c(a^{\dagger}a-(N+2))
                        a^{\dagger}&\hbar\omega a^{\dagger}a  
                        - \frac{\epsilon}{2}\cr}\right).
\ee
As it is clear from the Eq.(\ref{qes}), the Hamiltonian $H_T$ 
preserves the finite dimensional vector space namely
\be
{\cal V}_n = span\Biggl\{\left(\matrix{\mid j \rangle\cr
                        0\cr}\right),\left(\matrix{0\cr
                       \mid k\rangle\cr}\right)
     \ \ , \ \   j = N,\dots, 0 \ ; \ k = N+2,\dots, 0
                    \Biggr\},
\ee
and $n$ is fixed according to $n = N+2$.
From this, we   conclude that the
Hamiltonian $H_T$ is quasi-exactly solvable.
Hence the terms of perturbation added to $H_{12}$
have broken its non solvability.

Notice that is also easily to reveal the quasi-exact
solvability of the operator expressed in Eq.(\ref{operator})
by considering  the matrix Hamiltonian  Eq.(\ref{operator})
in terms of differential expressions. Here we have to replace
the operators $a^{\dagger}$ and $a$ respectively by their
differential expressions given by Eq.(\ref{creation}),
performing the standard gauge transformation as,
\be
\tilde H_T = exp({\frac{\omega x^2}{2}}) \  H_T  \ \ 
exp({-\frac{\omega x^2}{2}}),
\ee
 and thus, after some algebra, we obtain a matrix Hamiltonian 
 which preserves
 the finite dimensional vector space of the form ${\cal V}_k = (P_k(x), P_{k+2}(x))^t$
 with $k\in \N$ and $n = k+2$ ( i.e $n$ which is expressed 
 in Eq.(\ref{operator})).
 This operator $\tilde H_T$ (therefore $H_T$) is quasi-exactly 
 solvable because it
 is expressed in terms of the integer $n$  which is fixed according 
 to $n = k+2$.

\section{Spectral properties}
In this section, we would like to emphasize a few properties
of the spectrum of the Hamiltonian discussed above.
First we stress that for given $k$ the JC model admits $k$ levels
which are $\rho$-independant and which are not involved
in the list given above. They are of the form
$$\psi_j =
\left(\matrix{
\vec 0 \cr
\vert j \rangle \cr}\right)
 \ \  , \ \  0 \leq j \leq k-1 \ \ ,$$
where $\vec 0$ denotes the null vector of the Hilbert space.
The corresponding eigenvalue is $E_j = j- \frac{\epsilon}{2}$.

The spectrum of the JC model (and of its generalisations for $k > 1$)
varies considerably with the parameter $\rho$. In Fig. 1, we show
the evolution of six levels in the $k=2, \phi=1$ case.
They correspond to the two $\rho$-independant eigenstates
and the ones with $n=0,1$ in Eq.(34).
In Fig. 1 and in the following we assume $\epsilon = 1$
for simplicity but the features pointed out below remain similar for
$\epsilon \neq 1$. The same levels corresponding to the non hermitian
case $\phi=-1$ are reported on Fig. 2. The contrast with Fig.1 
is obvious. Couples of eigenvalues regularly disappear at finite
values of the coupling constants $\rho$. So that, at finite $\rho$
only a finite  number of real eigenvalues subsist, the other being
real. In this respect, the Hamiltonian is like a quasi exactly 
solvable operator.

The energy levels displayed on Fig.1 corresponds
to the six lowest ones in the limit $\rho = 0$.  The figure
clearly shows that they mix relatively quickly for increasing $\rho$ and
that, for instance, eigenvectors involving two or more quanta become
the ground state for $\rho \sim 1$.

We have studied the evolution of the spectrum when the QES-extension
of the model,
$H_{12} = \rho a^2 + \theta a (1 - \frac{1}{N+2} a^{\dagger} a)$
namely characterized by the new coupling constant $\theta$,
is progressivel switched on. Notice that the vector
$\psi_0 = (\vec 0 \ , \vert 0 \rangle)^t$ is an eigenvector
with $E = - \epsilon/2$,
irrespectively of $\rho, \theta$

In the case $\rho = 0, N=1$ the effect of the new term on the
eigenvalues under consideration leads to
$$
  E = - \frac{1}{2} \ , \ \ \frac{1}{6}(3 \pm 4 \theta)
  \ \ , \ \ \frac{1}{6}(9 \pm 2 \sqrt{2}) \theta ,   \ \ \frac{5}{2}
$$
These levels are indicated on Fig. 3 by the dotted lines and it is
clearly seen that they also lead to numerous level crossing.

The evolution of the eigenvalues corresponding to the case
$\rho = 1$ is displayed by the dashed lines in Fig.3, supplemented by
the black line $E = - 1/2$ which is present irrespectively of $\rho$.
The figure clearly shows that the occurence of the new term
induced only one level mixing,
namely two levels cross at $E = -1/2$ for $\theta = 1.5$
For larger values of $\rho$, e.g. $\rho = 2$,
the analysis reveals that the algebraic eigenvalues
depend only weakly of $\theta$.

\section{Series expansion and Recurence relations}
Here we would like to present another aspect of the QES Hamiltonian presented
in the previous section. Following the ideas of \cite{benderdunne} 
we will construct
the solution for energy $E$ under the form of a formal serie
in the basic vector
whose coefficients are polynomials in $E$.
More precisely, we write the solution of the equation
\be
H_T\psi = E\psi,
\ee
in the form
\be
\psi = \left(\matrix{
\sum_{j=0}^{\infty}p_j(E)\mid j\rangle\cr
\sum_{j=-2}^{\infty}q_j(E)\mid j+2\rangle\cr}\right)
\ee
and where $H_T$ is given by the Eq.(\ref{operator}).
After some algebra it can be seen that the polynomials $p_j(E), q_j(E)$ 
obey the following recurence relations
\be
A_{j+1}P_{j+1} + B_jP_j = 0,
\ee
where
\begin{eqnarray}
&A_{j+1} &=\left(\matrix{\rho\sqrt{(j+2)(j+3)}&-(E-(j+1)-\frac{\epsilon}{2})\cr
                       0&\hat c(j+2-n)\sqrt{j+2}\cr}\right),\nonumber\\
&B_j &=\left(\matrix{c(j+2-n)\sqrt{j+2}&0\cr-(E-(j+2)+\frac{\epsilon}{2})
                       &\rho\sqrt{(j+1)(j+2)}\cr}\right),\nonumber\\ 
&P_j &=\left(\matrix{q_j\cr 
p_j\cr}\right) \ , \ j=-2,-1,0,1,\dots                                                             
\end{eqnarray}
These equations have to be solved with the initial conditions
\be
q_{-2} = 0 \ , \ q_{-1} = {\cal N}
\ee
with ${\cal N}$ fixing the normalisation of the solution.
Then the solution for $q_j$ turns out to be a polynomial of degree $E^{2j}$.
The quasi-exact solvability of the system leads to the fact that 
$A_{n-1}$ is not
invertible and that $p_{n-1}$ can be choosen arbitrarily.
With the choice $p_{n-1} = 0$
it turns out that all polynomials $p_j, q_j$ with $j \geq n-2$
are proportional to $q_{n-3}(E)$.
As a consequence for fixed $n$ and for the values of $E$ 
such that $q_{n-3}(E) = 0$ the serie
above is truncated and the set of algebraic eigenvectors are recovered.
We would like to stress that series considered in this section are built
with the basis vector of the harmonic oscillator and not on monomials in
 $x$ contrasting with the construction of Ref.\cite{benderdunne}.
In the case of standard QES equations \cite{benderdunne} there it appears
 a three terms  recurence relations which leads to sets of orthogonal
 relation. In the case of systems of QES equations adressed in \cite{bbn}
 the recurence relation is also three terms but the situation here is quite
 different.
Actually, it is to our knowledge, an open question to know whether the set of
polynomials $(p_j(E), q_j(E))$ are somehow orthogonal
as it is the case for standard scalar equation.

\section{Hidden algebraic structures}
As pointed out in the previous sections, the different
Hamiltonians studied here posses the property that
their spectrum can be (partly or fully) computed.
This property is deeply related to the fact that
the corresponding operators are elements of the enveloping
algebra of particular graded algebra in an appropriate finite
dimensional representation.
The classification of linear operators preserving the vector
spaces ${\cal V}(m,n) = (P_m(x),P_n(x))^t$ 
was reported in \cite{bk}. It is shown that these operators
are the elements of the enveloping algebra of some non-linear
graded algebra depending essentially of $\vert m-n \vert$.
Note that, in the present context, the difference
$\vert m-n \vert$ is nothing else but  the parameter called $k$
in Sect. 3.
The cases $k=1$ and $k=2$ are special because the underlying
algebra is indeed a graded Lie algebra.
In the case $k=1$, related to the conventional JC model,
the Hamiltonian is an element of the enveloping algebra of
$osp(2,2)$; in the representation constucted in \cite{shiftur}.
 The generators involved in this relation  do not depend explicitely
 on $n$, i.e. of the dimension of the representation, explaining
 that the Hamiltonian is exactly solvable.
Finally, in the case $k=2$, the Hamiltonian is an element of the
graded Lie algebra q(2), as shown in \cite{dv,bh}.
 This algebra possesses an sl(2)$\times$U(1) bosonic subalgebra
and six fermionic operators splitted into three triplets of the
sl(2) subalgebra. In the case of the JC model corresponding to $k=2$, 
the Hamiltonian
is independant on the dimension of the representation $n$ and
the model is exactly solvable. For the modified model of Sect. 4, the
supplementary interaction term $H_I$ defined in (40) indeed depends
on $n$ and the operator admit only the vector space ${\cal V}_n$
as finite dimensional invariant vector space.
\section{Conclusions}
In this letter, we have considered several extensions of JCM by
adding to its original Hamiltonian the polynomial $P(a^{\dagger}a)$
of degree $d \geq 2$ and an arbitrary sign, say $\phi=\pm 1$,
 in the non-diagonal
 interaction term.
In fact, considering this  sign $\phi=-1$, these extended Hamiltonians
are nonhermitian and not PT invariant but they satisfy the
pseudo-hermiticity with respect of different operators $P$
and $\sigma_3$. This new property reveals the reality of the
energy spectrum which has been constructed algebraically.
They become hermitian when one  considers the sign $\phi=1$.
Notice that these Hamiltonians are completely solvable as
it has been pointed out by the QES technique.

Several usual properties available with hermitian Hamiltonian
are not kep with pseudo-hermitican.
Namely the eigenstates given by (\ref{state0})
and (\ref{state1})(i.e corresponding to the doublet
$\mid 0,\frac{1}{2}\rangle$ and $\mid k,-\frac{1}{2}\rangle$) are
not orthogonal to each other, but they are orthogonal to all
eigenstates corresponding to other doublets.
For example, the eigenstate (\ref{state0}) and
the one given by Eq.(\ref{cos})(i.e it corresponds
to the doublet $\mid n, \frac{1}{2}\rangle$ and $\mid n+k, -\frac{1}{2}\rangle$)
are orthogonal to each other.
The eigenstates of any particular doublet are orthogonal
to each other only if $\theta_m = m\pi$
(i.e with $m = 0,1,2,\dots,k,\dots,n+k$),
this implies $\rho = 0$ because it depends
to $\sin\theta_m$. In fact, as the energy eigenvalues
are entirely real, it is impossible to have all
eigenstates orthogonal to each other.
This is explained by the unbroken symmetry of the
operator $P\sigma_3$. But for energy eigenvalues complex,
the orthonormality condition is satisfied by all the associated
eigenstates. All these discussions are the result of the
scalar product applied to those eigenstates.

We manage to construct a JC-type Hamiltonian describing both
one and two-photons interactions in terms of quasi exactly solvable
operators. This involves a very specific interaction term
of degree one in the creators and annihilators which can be seen
as a perturbation of more conventional p-photons interacting term.
Several properties of this new family of QES-operators have been
presented. Namely, (i) they can be written in terms of the generators of
the graded Lie algebra osp(2,2) in a suitable representation; 
(ii) when expressed as  series, the formal solutions of $H_T \psi = E \psi$
leads to a different type of recurence relation between the different
terms of the series.
\newpage
\begin{figure}
\epsfysize=22cm
\epsffile{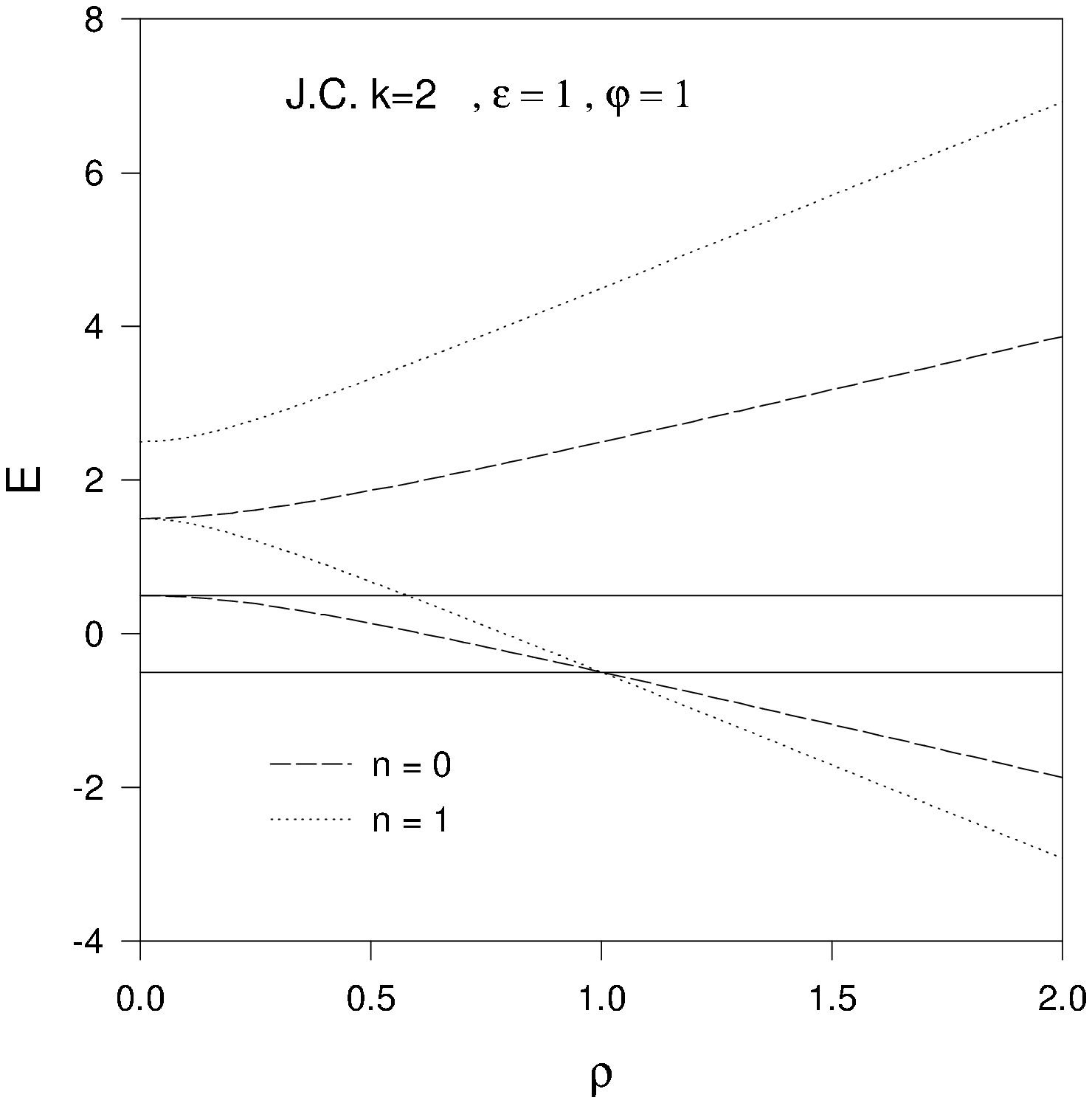}
\vskip -3cm
\caption{\label{Fig.1} The first few energy levels in the
$k=2$-JC Hamiltonian for   $\epsilon = 1$ and $\phi = 1$.
}
\end{figure}
\begin{figure}
\epsfysize=22cm
\epsffile{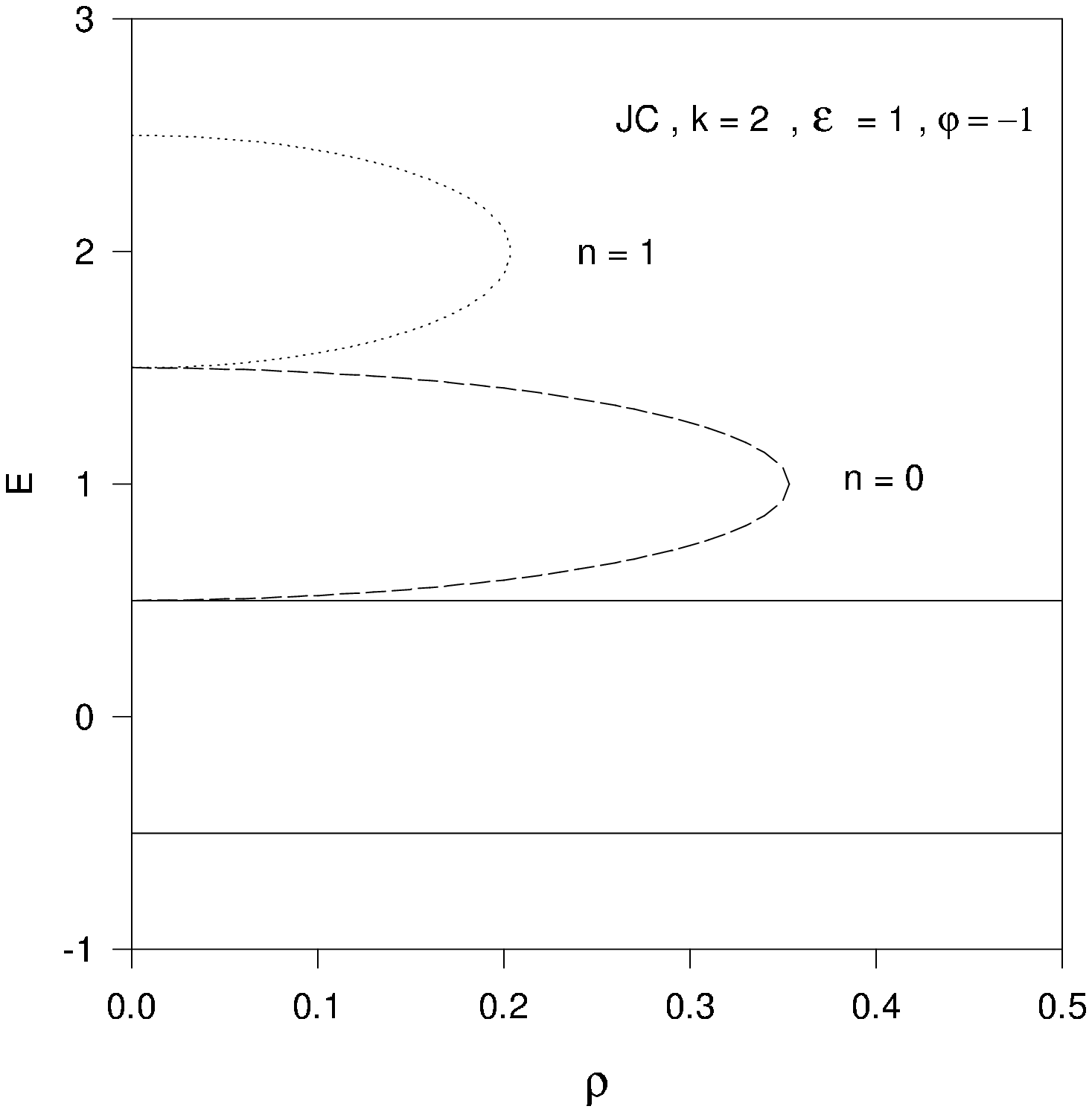}
\vskip -3cm
\caption{\label{Fig.2} The first few energy levels in the
$k=2$-JC Hamiltonian for        $\epsilon = 1$ and $\phi = -1$.
}
\end{figure}
\begin{figure}
\epsfysize=22cm
\epsffile{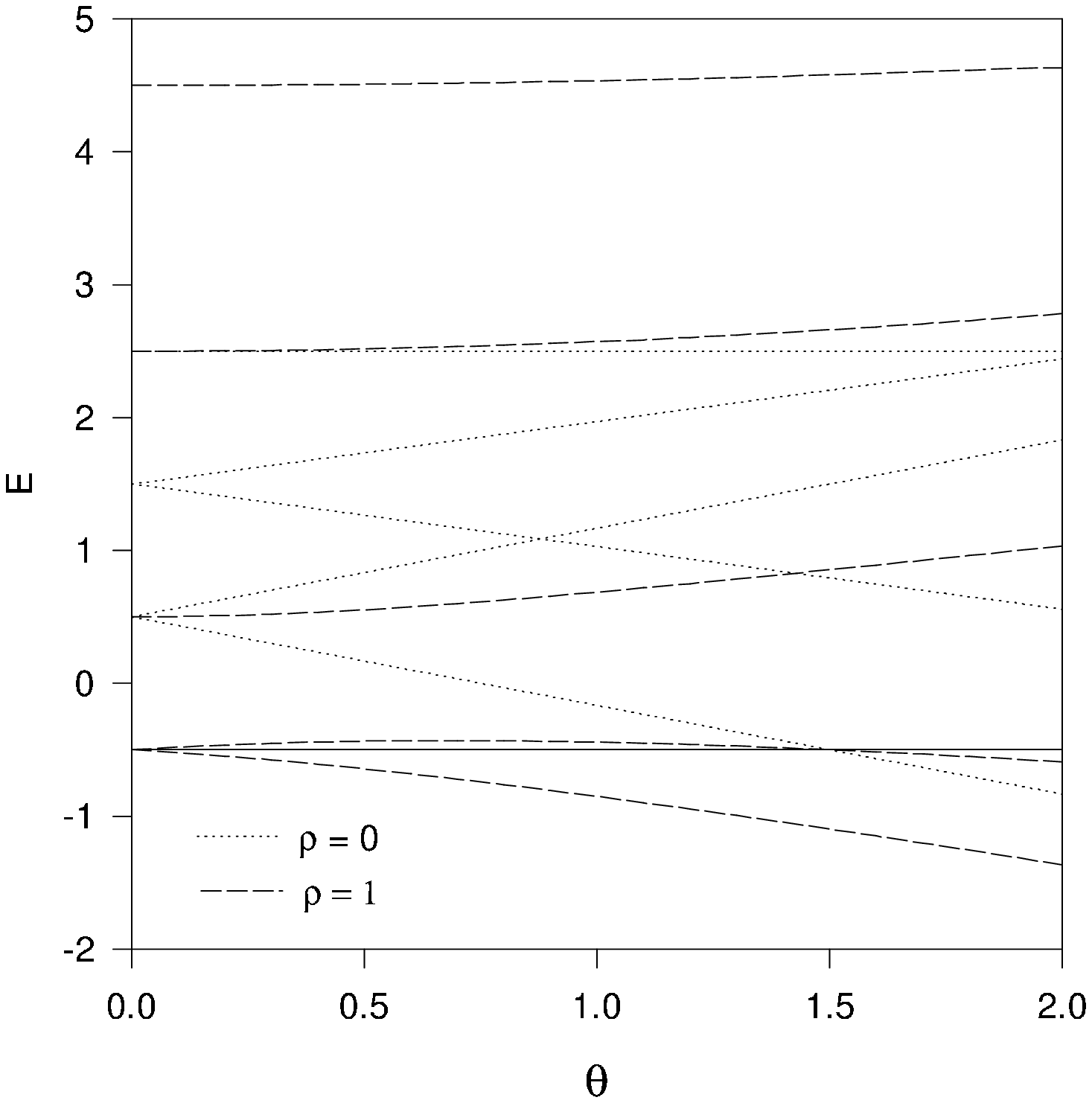}
\vskip -3cm
\caption{\label{Fig.3} The first few energy levels in the QES
deformed $k=2$ JC Hamiltonian as function of the
parameter $\theta$, the energy level $E=-1/2$ (solid line) is
independant of $\rho$.
}
\end{figure}


\newpage
\begin{thebibliography}{99}  
\bibitem{bender1} C. M. Bender and S. Boettcher, Phys. Rev. Lett. {\bf{80}}, 5243(1998); C. M. Bender, S.
 Boettcher and P. N. Meisinger, J. Math. Phys. {\bf{40}}, 2210(1999).
\bibitem{bender2} C. M. Bender and S. Boettcher, Phys. Rev. Lett. {\bf{89}}, 270401(2002).
\bibitem{bender3}C. M. Bender hep-th/0501052.
\bibitem{mostafa} A. Mostafazadeh, J. Math. Phys. {\bf{43}}, 205(2002);  {\bf{43}}, 2814(2002);  {\bf{43}},
 3944(2002); A. Mostafazadeh and A. Batal, J. Phys. {\bf{A37}}, 11645(2004).
\bibitem{Mandal} B. P. Mandal, hep-th/0412160
Mod. Phys. Lett {\bf A 20} 655 (2005).
\bibitem{Pijush} P. K. Ghosh,
quant-ph/0501087, J. Phys. {\bf A 38}, 7313 (2005).
\bibitem{turbiner} A. Turbiner, Commun. Math. Phys. {\bf{119}}, 467 (1988).
\bibitem{ushveridze} A.G. Ushveridze, {\it Quasi exactly Solvable Models in
Quantum Mechanics} (IOP 1995).
\bibitem{shiftur} M. Shifman and A. Turbiner, Commun. Math. Phys.
\bibitem{bk} Y. Brihaye and P. Kosinski,  J. Math. Phys. {\bf 35}
3089 (1994).
\bibitem{Gerry} C. C. Gerry, Phys. Rev. {\bf{A37}}, 2683(1988).
\bibitem{Ray} B. Deb and D. S. Ray, Phys. Rev.{\bf{A48}}, 3191(1993).
\bibitem{knight} P. L. Knight and P. M. Radmore, Phys. Rev. {\bf{A26}}, 
676(1982).
\bibitem{nathalie} N. Debergh and A.B. Klimov, J.Mod.Phys.{\bf {Vol.16}}, 4057(2001).
\bibitem{benderdunne} C. M. Bender and G. V. Dunne, J.Math. Phys. {\bf{37}}, (1996).
\bibitem{bbn} Y. Brihaye, J. Ndimubandi and B. Prasad Mandal, "QES polynomials,
invariant spaces and polynomials recursion", math-ph/0601004.

\bibitem{dv} N. Debergh and J. Van der Jeugt, J. Phys. {\bf A 34},
\bibitem{bh} Y. Brihaye and B. Hartmann, Phys. Lett. {\bf A 306}, 291
\end{thebibliography}
\end{document}